\documentclass{aa}
\usepackage{graphicx,natbib}
\bibpunct{(}{)}{;}{a}{}{,}

\begin{document}

\sloppypar

   \title{Quenching of the accretion disk strong aperiodic variability at the
magnetospheric boundary}

   \author{M. Revnivtsev \inst{1,2}, E. Churazov \inst{3,2}, K. Postnov \inst{4}, S. Tsygankov \inst{3,2}}

%   \offprints{revnivtsev@iki.rssi.ru}

   \institute{
	Excellence Cluster Universe, Technische Universit\"at M\"unchen, Boltzmannstr.2, 85748 Garching, Germany
\and
              Space Research Institute, Russian Academy of Sciences,
              Profsoyuznaya 84/32, 117997 Moscow, Russia
\and
	Max-Planck-Institut fuer Astrophysics, Karl-Schwarzschild-str.1 , 85741, Garching, Germany
\and
	Sternberg Astronomical Institute, Moscow State University, Universitetskij pr., 13, 119899, Moscow, Russia
            }
  \date{}
\authorrunning{Revnivtsev et al.}
\titlerunning{Quenching of the accretion disk strong aperiodic variability}

\abstract{We study power density spectra (PDS) of X-ray flux
  variability in binary systems where the accretion flow is
  truncated by the magnetosphere.
  PDS of accreting X-ray pulsars where the neutron star is close to the
  corotation with the accretion disk at the magnetospheric boundary,
have a distinct break/cutoff at the neutron star spin
  frequency. This break can naturally be explained in the
  ``perturbation propagation'' model, which assumes that at any given
  radius in the accretion disk stochastic perturbations are introduced
  to the flow with frequencies characteristic for this radius. These
  perturbations are then advected to the region of main
  energy release leading to a self-similar variability of X-ray flux  $P\propto
  f^{-1...-1.5}$. The break in the PDS is then a natural manifestation of
  the transition from the disk to magnetospheric flow at the
  frequency characteristic for the accretion disk truncation radius
  (magnetospheric radius). The proximity of the PDS break frequency to
  the spin frequency in corotating pulsars strongly suggests that the typical
  variability time scale in accretion disks is close
to the Keplerian
  one. In transient accreting X-ray pulsars characterized by large
  variations of the mass accretion rate during outbursts, the
  PDS break frequency follows the variations of the X-ray flux,
  reflecting the change of the magnetosphere size with the accretion
  rate. Above the break frequency the PDS steepens to $\sim f^{-2}$
  law which holds over a broad frequency range. These results suggest
  that strong $f^{-1...-1.5}$ aperiodic variability which is ubiquitous in
accretion disks is not characteristic for magnetospheric flows.
 }

   \maketitle

%
%________________________________________________________________

\section{Introduction}

It has been recognized
since the beginning of X-ray astronomy that the
flux of accreting X-ray binaries demonstrates strong aperiodic variability \cite[see
e.g.][]{rappaport71,oda74}.  Almost immediately after the discovery,
the noise in the X-ray light curves of accreting binaries (like e.g. Cyg
X-1) was explained as a superposition of randomly occurring
X-ray emission flashes (shots) of similar duration (the shot noise model, \citealt{terrell72}). This provided an explanation to the shape of
the power density spectra (PDS, the Fourier transform of the
autocorrelation function of the lightcurve of a source) of different
X-ray sources.

However, the accumulation of more data posed serious questions to this
paradigm. In particular, it was very hard to explain a huge range of
the X-ray variability time scales observed in some sources \cite[see
e.g.][]{churazov01} and the linear correlation between the variability
amplitude and the average flux of sources \citep{uttley01}.
Indeed, to explain observed large variability amplitude, the individual flashes/shots in the shot noise model should be very powerful. Therefore these flashes must come from the innermost
region of the accretion flow, where most of the energy is released. The characteristic time scales in this region are
very short -- milliseconds or tens of milliseconds for stellar-mass
compact objects. However, very often, e.g.  in the soft/high state of
accreting binaries, the observed power spectra have a
power law shape extending down to frequencies as low as
$10^{-5}-10^{-6}$~Hz, i.e. 5-7 orders of magnitude longer time scales than all
time scales characteristic for the region of the main energy release \cite[see][]{churazov01,gilfanov05}.

A very promising model for the aperiodic X-ray
variability of accreting sources is the "perturbation propagation" model \citep{lyubarskii97,churazov01,kotov01,arevalo06}. In this model, the
X-ray flux variability is caused by
the variations of the instantaneous value of the mass accretion rate
in the inner accretion flow.
In turn, the variations of the mass accretion rate are due to the perturbations introduced to the accretion flow by the stochastic variations of the disk viscous stresses.
In this model the observed variability is a multiplicative superposition of perturbations introduced at different radii.
Assuming that the fractional amplitudes of the mass accretion rate perturbations are the same at all radii,
the PDS of the emerging lightcurve will naturally
appear as a self-similar power-law with slope $-1...-1.5$ up to the maximal
frequencies that can be generated in the disk \citep{lyubarskii97}.
Direct magneto hydrodynamic simulations of accretion flows
\cite[se e.g.][]{brandenburg95,balbus99,hirose06} provide further support to
this semi-phenomenological model. In particular, these simulations
show that perturbations in the instantaneous mass accretion rate generated at any given radius of the disk have
characteristic time scale proportional to the local dynamical time.

This model of the aperiodic X-ray flux
variability implies that the presence of the accretion disk
edges, both outer and inner, should be reflected in
the noise properties of the X-ray light curve.
Signatures of outer edges
of accretion disks were found by \cite{gilfanov05}
in the low frequency parts of the noise power spectra
of low mass X-ray binaries.

Accretion disks around compact stars in X-ray binaries should also have inner edges,
which should manifest itself in
the power spectrum of their X-ray light curves. Specifically,
a definite break is expected to be present in the PDS
at the characteristic frequency of variability
generated at the inner edge of the disk.
At frequencies below this break the power spectrum is expected to
be produced in the accretion disk and have a self-similar
slope about -1.0...-1.5 \citep{lyubarskii97,churazov01,gilfanov05,revnivtsev06}, while at
higher frequencies the character of the flow changes and
the PDS slope may be different.

In accreting X-ray pulsars and intermediate polars
the central compact object (a neutron star or a white dwarf)
has a strong magnetic field which can disrupt the disk-like accretion flow
at the magnetospheric boundary, or even prevent
the formation of accretion disk at all (like in polars),
dividing the flow into two distinct parts -- the accretion disk
and the magnetospheric flow. The noise properties of these flows
may be very different.

In the present paper we compare PDS of different types of accreting
X-ray binaries and discuss the observational support for the qualitative picture of aperiodic variability outlined above.

\section{Truncated accretion disks in different classes of sources}

We shall consider several classes of accretion X-ray binaries:

\begin{itemize}

\item X-ray binaries with compact objects, which have
  magnetospheres powerful enough to disrupt the accretion disk at
  large distances (accreting X-ray pulsars, intermediate polars)

\item X-ray binaries with large magnetospheres in which the
  accretion disk does not form and the accretion proceeds along the
  magnetosphere from the very beginning (polars)
\end{itemize}

% In addition, we will make use of important property of accretion powered X-ray
% sources.
In a fair fraction
of persistent accreting X-ray pulsars the spin period of neutron star is observed to be close to synchronization (corotation) with the Keplerian rotation of the accretion
disk at the magnetospheric boundary,
which is explained by the standard description of the interaction of
accretion disk with a magnetized neutron star
\cite[see e.g.][]{davidson73,shakura75,lipunov76,gl79,corbet84,ziolkowski85}.

On the contrary,
due to much larger moments of inertia of white dwarfs, it takes a
much longer time to bring
accreting
magnetic white dwarfs in cataclysmic variables
into corotation with surrounding accretion disks,
% are typically far from corotation with the surrounding accretion disks and thus
and their magnetospheres rotate with periods much larger
than the Keplerian ones at the inner disk edge.

The perturbation propagation model makes distinct predictions for the noise properties of
accretion flows in
these classes of objects. Namely:

\begin{enumerate}

\item The noise power spectra of accreting sources with large magnetospheres
and without accretion disk (polars, e.g. AM Her), in which matter is
transferred directly from the companion star to the compact star
via magnetospheric accretion, should be different from those of sources with
accretion disks and without magnetospheres
(e.g. accreting black holes like Cyg X-1 in their soft state).

\item The noise power spectra of sources with accretion disk truncated at the
  magnetospheric boundary should have a break
  corresponding to the characteristic frequency in the disk near its inner
  boundary.

\item If the size of the magnetosphere changes (e.g. as a response to
the change in mass accretion rate like in transient
  pulsars), the break frequency must change correspondingly.

\item For persistent X-ray pulsars which are close to corotation,
  the comparison of the PDS break frequency with the compact object
  spin frequency can be used to determine the ratio of the
  characteristic frequency of perturbations generated in the disk to
  the Keplerian frequency.
\end{enumerate}

\begin{figure}[htb]
\includegraphics[width=\columnwidth]{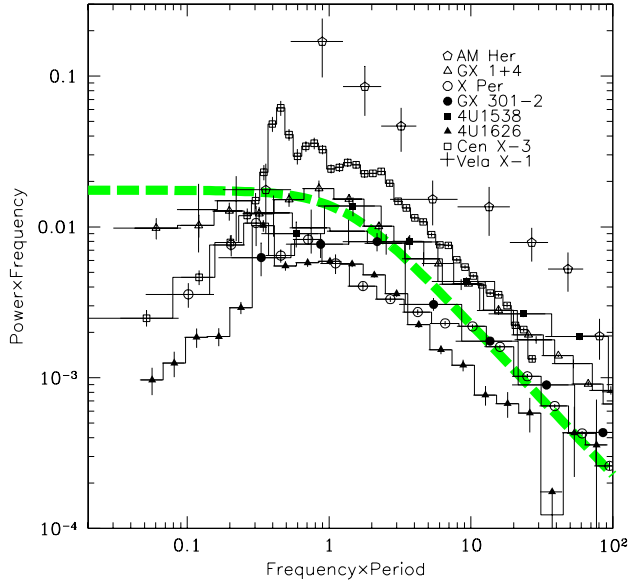}
\caption{Power density spectra of persistent accretion powered X-ray pulsars. The frequency shown along the X-axis is expressed in units of the compact object spin frequency. The power density spectra are multiplied by frequency to show the square of the fractional RMS per decade of frequency.  For comparison with the observed profiles, the thick dashed curve
shows an analytical model $P\propto f^{-1}\,(1+(f/f_0)^2)^{-0.5}$.}
\label{powers}
\end{figure}

\subsection{Breaks in the power spectra spectra of magnetized accretors}

It has been noticed in earlier studies \cite[e.g.][]{hoshino93}
that noise power spectra of accreting X-ray pulsars
typically have breaks around the pulse frequency. This is shown
more clearly in Fig.\ref{powers}, where we plot the noise power spectra
of several persistent X-ray pulsars, including
Cen X-3, 4U 1626-67, Vela X-1, 4U 1538-52, GX 301-2, X Persei, GX 1+4, and of the
magnetic white dwarf (polar) AM Her. The frequency scale of their power spectra is normalized
through multiplication by the pulse period of the sources. The variability, associated with regular pulsation
was removed from the original X-ray light curves
by subtracting folded segments of light curves
with a duration of 10-20 spin periods .
Strictly speaking this procedure does not remove the contribution of regular pulsations completely since often the periodic and aperiodic variabilities show signs of nonlinear interactions (see e.g. the discussion in \citealt{tsygankov07}); however, it is good enough as the first approximation (see e.g. \citealt{finger96}).

It is seen that {\sl all power
spectra of accreting X-ray pulsars show clear breaks approximately
at their spin frequency}. PDSs of all sources have a similar power-law slope
above the break frequency irrespective of the PDS form below it.

In the framework of the perturbation propagation model
(see e.g. \citealt{lyubarskii97,churazov01}), we can interpret
this observational fact as a signature of the truncation
of the accretion disk flow and its conversion into a magnetospheric flow.

The proximity of the PDS break frequency to the spin frequency
(which in the case of corotating systems is close to the frequency of the Keplerian rotation around the compact object at the
inner edge of the accretion disk) allows us to conclude that
{\sl the characteristic time scale of variability
produced at some distance from the central
compact objects are close to the local Keplerian time scale.}

Fig. \ref{powers} shows that above the break frequency the power
spectrum of the flux variability typically is a power law with the slope
close to $\sim -2$. We can not exclude that such slope  of the flux
variability is a property of the magnetospheric accretion, however,
we note that similar slopes of power spectra
sometimes observed in accreting systems without magnetospheric accretion,
for example unmagnetized cataclysmic variables
\cite[see e.g.][]{kraicheva99,pandel03}.

\subsection{Change of the magnetospheric size with mass accretion rate}

% If our interpretation of the shape of PDSs of accreting pulsars is correct
If the break frequency in the noise power spectra of accreting X-ray pulsars
indeed reflects the time scale of the noise generation at the
inner boundary of the accretion disk/flow, its value should
depend on the mass accretion rate in the binary system. Increase in the mass accretion
rate decreases the size of the magnetosphere (and hence the inner radius of the disk) and
%%%%%%%%%%%%%%%%%%%%%%%%%%%%%%%%
brings the system off the corotation, so
the characteristic frequency at the inner edge of the disk/flow increase.
A similar situation should be observed in the case of luminous intermediate polars (accreting
magnetized white dwarfs with moderate magnetospheres) which typically are always out of the corotation.

We can verify this hypothesis by examining power spectra of bright transient X-ray pulsars (like
A0535+26, 4U0115+63, V0332+53, KS 1947+300), which demonstrate wide range of X-ray luminosities during the outbursts, and power spectra of luminous intermediate polars (e.g. V1223 Sgr).

\begin{figure}[htb]
\includegraphics[width=\columnwidth,bb=20 150 600 600,clip]{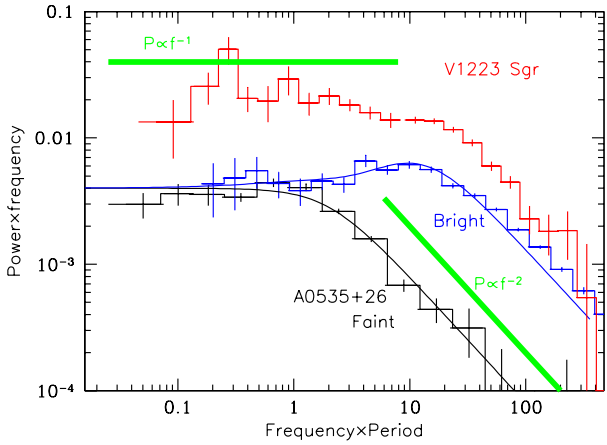}
\caption{Power density spectra of accreting X-ray pulsar A0535+26 at low accretion rate (labeled as "faint") and in the strong spin-up regime during the outburst (labeled as "bright"). The upper curves shows the power spectrum of luminous intermediate polar V1223 Sgr}
\label{dif0535}
\end{figure}

In Fig.\ref{dif0535} we show examples of power spectra of the pulsar
A0535+26 during its bright outburst in 2005 and the power spectrum of V1223 Sgr.
The power spectra of A0535+26 were averaged over two time intervals:
a) with small X-ray luminosity ($L_{\rm x}\sim10^{36}$ erg/s) (low mass accretion rate)
and b) with higher luminosity ($L_{\rm x}\sim10^{37}$ erg/s) (higher mass accretion rate). The power spectrum of V1223 Sgr is averaged over all publicly available RXTE observations.

From Fig.\ref{dif0535} it is clear that the power spectra of A0535+26 in different luminosity states
differ significantly
at the frequencies higher than the pulse frequency, while at lower frequencies the PDSs are
almost identical. The difference at high frequencies
is essentially an addition of an extra noise component, which is presumably generated in the ring of accretion disk between the radii corresponding to the magnetosphere size in the high accretion rate (small radius) and low accretion rate (large radius). This ring and associated variability were absent in the state with low accretion rate.

The noise power spectrum of V1223 Sgr closely resembles that of A0535+26 in the bright state.
This similarity reflects the fact that both systems are out of corotation with their accretion disks -- the magnetospheres are squeezed by the increased accretion flow, and the inner parts of the accretion disks rotate much faster than the central object.

If the characteristic frequency $f_0$ of the noise at the magnetospheric boundary $R_{\rm m}$ is proportional to the frequency of the Keplerian rotation $\nu_{\rm K}$ of matter at the inner edge of the accretion disk $R_{\rm in}\approx R_{\rm m}$, we can relate the observed break frequency to the instantaneous value of the mass accretion rate $\dot{M}$ (see e.g. \citealt{pringle72,lamb73,davidson73,bildsten97}):
$$
2\pi \nu_{\rm K} = (GM)^{1/2} R_{\rm m}^{-3/2}
$$
where
$$
R_{\rm m}\approx \mu^{4/7} (GM)^{-1/7} \dot{M}^{-2/7}
$$
is the standard expression for the magnetospheric radius,  $\mu$
is the dipole magnetic moment of the neutron star.
Therefore we can anticipate that the break frequency will follow the dependence:
\begin{equation}
f_{\rm b}\propto\nu_{\rm K}\propto (GM)^{10/14} \mu^{-6/7} \dot{M}^{3/7}
\label{fb}
\end{equation}

\begin{figure}[htb]
\includegraphics[width=\columnwidth]{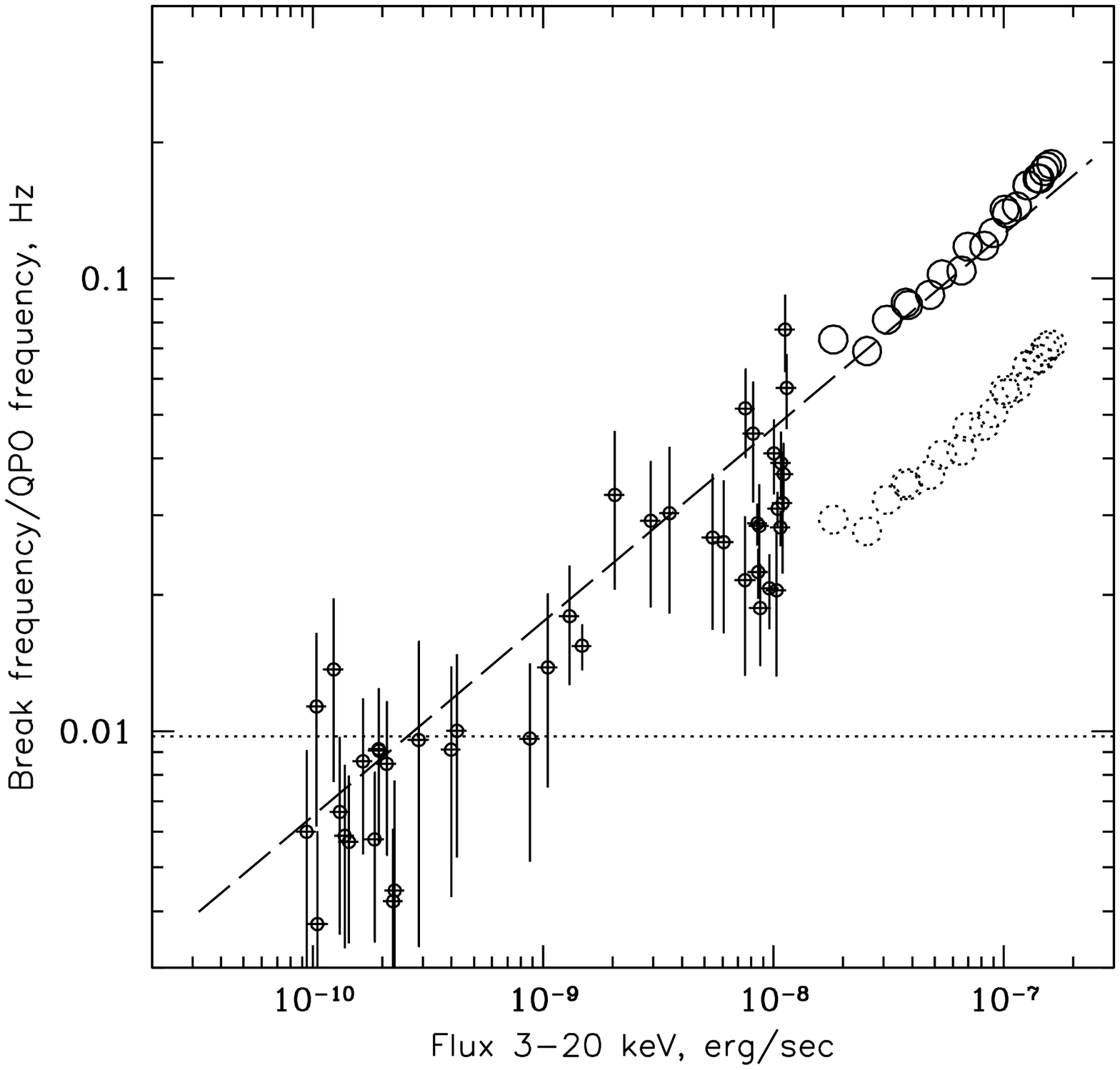}
\caption{Dependence of the break frequency in the noise power spectrum
of A0535+26 on the 3-20 keV X-ray flux (filled circles).
The dependence of the QPO frequency observed by \cite{finger96}
during the source outburst in 1994 is shown by open dotted circles. The solid circles show the results of \cite{finger96} when the QPO frequency is recalculated to the break frequency as $f_{break}=2.5\times f_{QPO}$
(see Fig. \ref{supportingfigures}, the upper panel). The dashed line shows the prediction of the
simplest ``magnetospheric'' model of the break frequency $f_{\rm b}\propto {L_{\rm x}}^{3/7}$ described in the text.
The dotted line shows the neutron star spin frequency.}
\label{correlation}
\end{figure}

It is exactly what we see during the evolution of the outburst of A0535+262
observed by RXTE in 2005 (we have substituted the X-ray luminosity $L_{\rm x}\simeq
0.1 \dot M c^2$ instead of the
mass accretion rate $\dot{M}$ into Eq. (\ref{fb})).
The dependence of the break frequency ($f_b$ in Eq. (\ref{fb})) on the X-ray flux of A0535+262 is shown in Fig.~\ref{correlation} (filled circles). For this this plot
we have used all observations of RXTE of this outburst after $\sim$MJD 53613, when the stable regime of accretion was established (see \citealt{caballero08,postnov08}).

\begin{figure}[htb]
\vbox{
\includegraphics[width=\columnwidth]{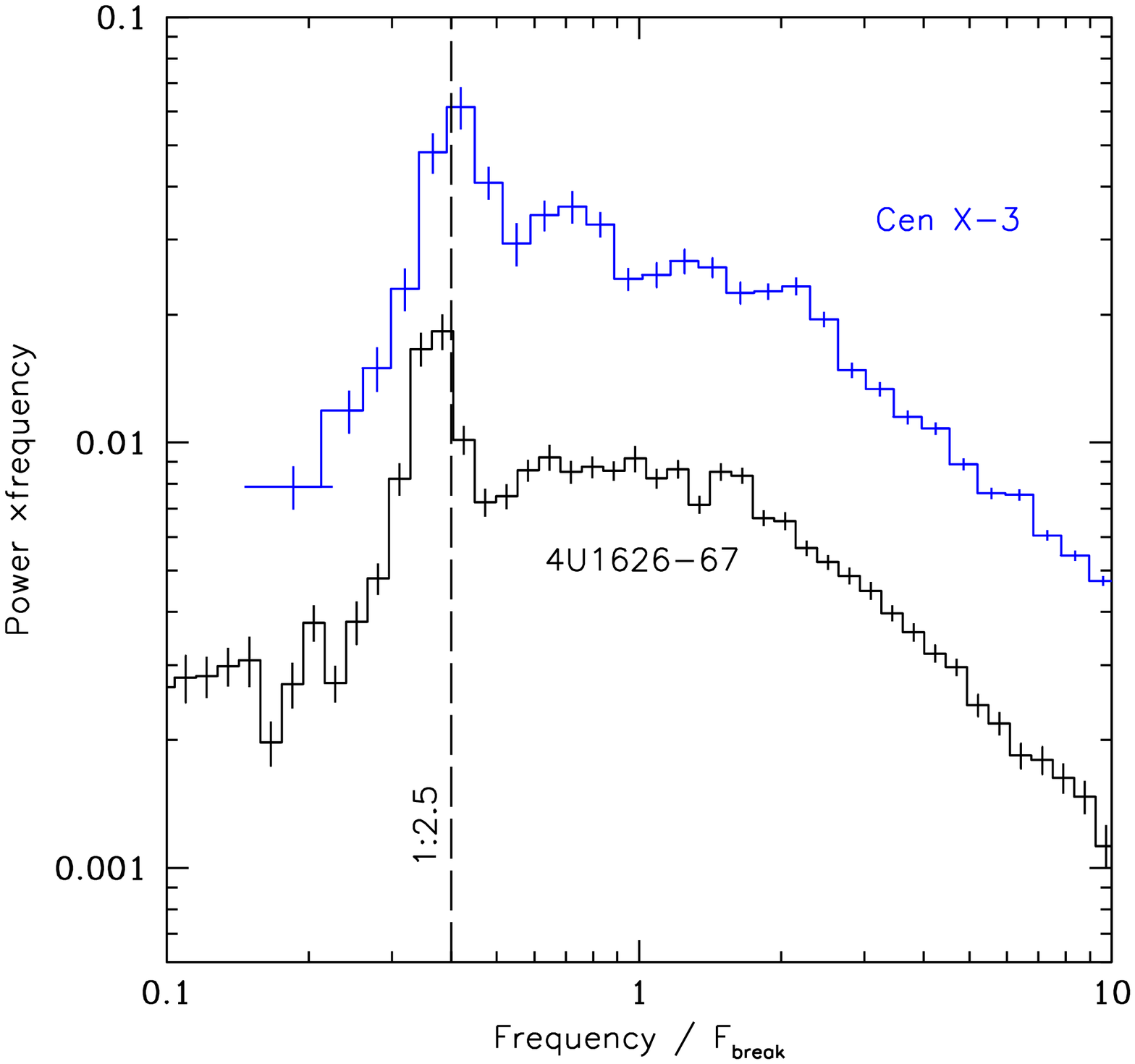}
\includegraphics[width=\columnwidth,bb=40 187 570 500,clip]{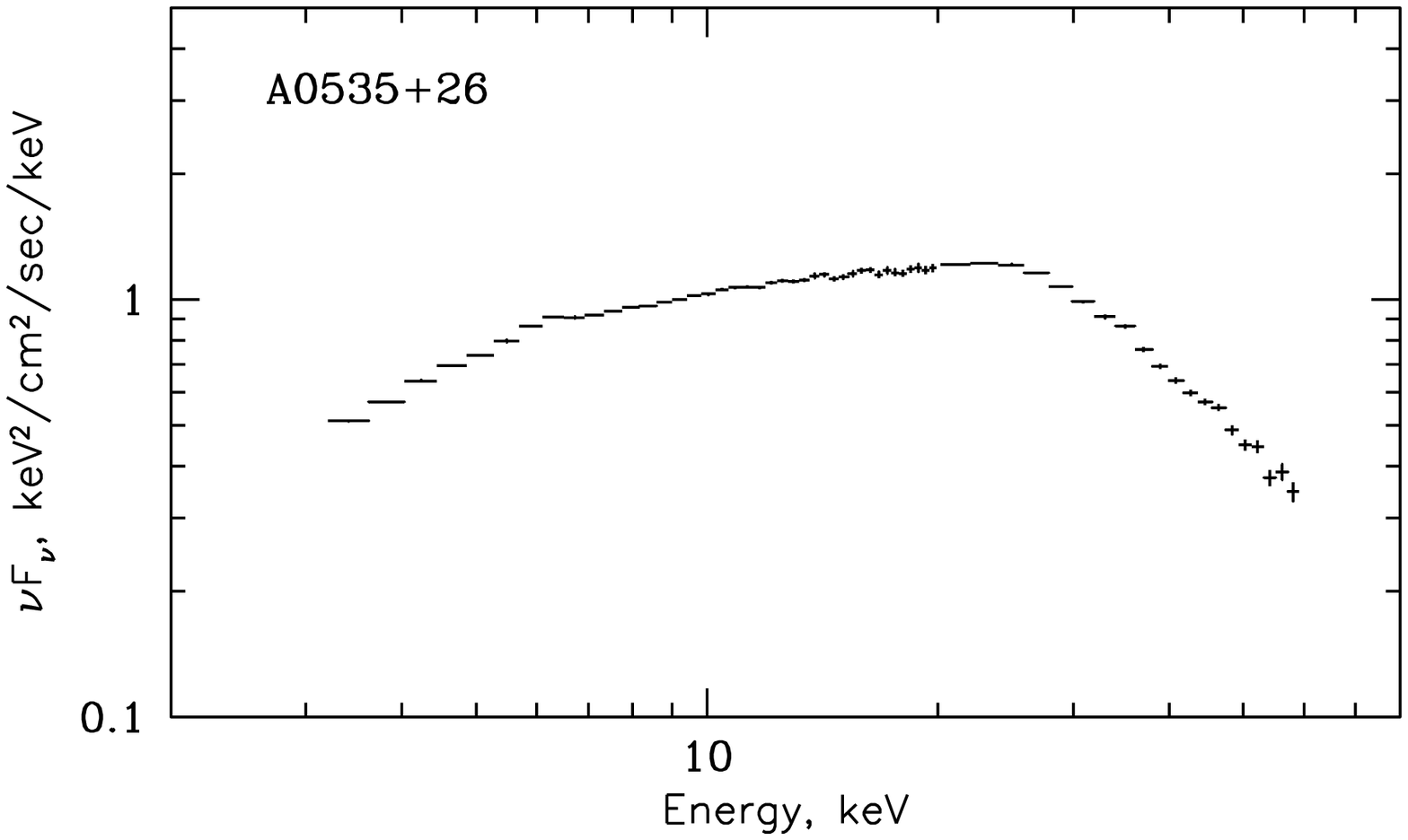}
}
\caption{Upper panel: power spectra of Cen X-3 and 4U1626-67 with low-frequency QPOs (the variability associated with regular pulsations has been  subtracted from the original X-ray light curves). The frequency axis has been scaled by the break frequency in the corresponding power spectrum. Lower panel: the broadband X-ray energy spectrum of A0535+26 observed by RXTE during its outburst in 2005. For this spectrum the ratio of energy fluxes
in 20-100 keV and 3-20 keV energy band is approximately $\sim$0.55.}
\label{supportingfigures}
\end{figure}

It is interesting to note that the dependence $f\propto L_{\rm x}^{3/7}$ was previously
established for the centroid frequency of quasi-periodic X-ray oscillations
in the power spectra of A0535+26 detected during its giant outburst
in 1994 \citep{finger96}. The authors argued that the oscillations originate at the inner
boundary of the accretion disk and are related either to the Keplerian rotation
at the inner edge of the disk or to the beat frequency between the Keplerian rotation and that of the neutron star magnetosphere \cite[e.g.][]{alpar85}.

Actually we can try to combine our present
RXTE measurements with QPO studies by \cite{finger96}
making use of two facts:
 \begin{itemize}
 \item[a)] when low-frequency QPOs are observed in power spectra
of accreting pulsars, their centroid frequency is related to the
break frequency as $\sim$1:2.5 (Fig.\ref{supportingfigures},
upper panel, see also \cite{angelini89} for QPO in EXO 2030+375). The correlation of the
QPO and the break frequencies is also observed in LMXBs, although
in these objects the centroid of the QPO feature is usually above
the break frequency \citep[see e.g.][]{wijnands99}.
\item[b)] during RXTE observations of the source outburst in 2005, the
X-ray flux of the A0535+26 in the energy band 20-100 keV (the range where measurements by \citealt{finger96} were done) is approximately factor of 0.55 of the X-ray flux in the energy band 3-20 keV (see the RXTE spectrum shown in the lower panel of Fig.\ref{supportingfigures}).
\end{itemize}
The dependence of the QPO frequency on the X-ray flux during its 1994 outburst, renormalized using the above factors, is shown in Fig.~\ref{correlation} (solid open circles). The renormalized dependence (solid open circles)
perfectly continues the
observed break frequency-flux dependence (filled circles) and corresponds to Eq. (\ref{fb}) (the dashed line).

Prominent QPO features similar to those detected by \cite{finger96} are not always observed in the power spectra of accreting X-ray pulsars. On the other hand a break in the PDS is more ubiquitous and therefore the diagnostics of the accretion flow based on the break frequency can be applied to larger datasets. For example, the break frequency
in the noise power spectrum can be used as an estimate of the dipole magnetic moment of compact objects using Eq. (\ref{fb}).

\section{Conclusions}
We studied aperiodic variability of the X-ray flux from accreting binaries,
in which the truncation of the disk-like accretion flow by the magnetosphere of
the compact object is important. The results can be
summarized as follows:

\begin{itemize}
\item There is a distinct break in the Power Density Spectra of
  accreting magnetized neutron stars and white dwarfs, apparently
  associated with the change of the disk-like accretion flow to the
  magnetospheric flow near the Alfvenic surface.

\item In transient systems with variable X-ray luminosity
the PDS break frequency $f_b$ changes with the X-ray luminosity (mass accretion rate) as $f_b\propto \L_x^{3/7}$,
in agreement with the standard theory of accretion onto magnetized compact stars.

\item This break can naturally be explained in the ``perturbation
  propagation'' model, which assumes that at any given radius in the
  accretion disk stochastic perturbations at frequencies
  characteristic for this radius are introduced to the flow. These
  perturbations are then advected by the flow to the region of main
  energy release leading to a self-similar form of the PDS
  $P\propto f^{-1...-1.5}$. The break in the PDS corresponds to a
  frequency characteristic for the accretion disk truncation radius
  (the magnetospheric radius).

\item We suggest that the PDS break frequency is directly related to the
  magnetospheric radius for a given value of the mass accretion rate
and can be used to estimate the magnetic moment
  of accreting compact stars.

\item For systems which are close to corotation (accreting X-ray pulsars) the
  PDS break frequency is close to the spin frequency of the neutron
  star. This strongly suggests that the characteristic frequency of
  perturbations introduced to the accretion flow in the disk
  is of order of the local Keplerian frequency.

\item In all studied objects the PDS above the break frequency follows the $P\sim f^{-2}$ law over a broad range of frequencies, suggesting that strong $f^{-1...-1.5}$ aperiodic variability which is ubiquitous in
accretion disks is not characteristic for magnetospheric flows.
\end{itemize}

\begin{acknowledgements}
The authors thank the anonymous referee for useful comments.
Authors thank Marat Gilfanov for useful discussions.
This research made use of data obtained from the High Energy
Astrophysics Science Archive Research Center Online Service,
provided by the NASA/Goddard Space Flight Center.
KP thanks the MPA for hospitality.
This work was
supported by DFG-Schwerpunktprogramme (SPP 1177), grants of Russian
Foundation of Basic Research 07-02-01051, 07-02-00961-a, 08-08-13734, NSh-5579.2008.2 and
the RAS program ``The origin and evolution of stars and
galaxies'' (P04).

\end{acknowledgements}


\begin{thebibliography}{}
\bibitem[\protect\citeauthoryear{Alpar
\& Shaham}{1985}]{alpar85} Alpar M.~A., Shaham J., 1985, Natur, 316, 239
\bibitem[\protect\citeauthoryear{Angelini, Stella,
\& Parmar}{1989}]{angelini89} Angelini L., Stella L., Parmar A.~N., 1989, ApJ, 346, 906
\bibitem[\protect\citeauthoryear{Ar{\'e}valo
\& Uttley}{2006}]{arevalo06} Ar{\'e}valo P., Uttley P., 2006, MNRAS, 367, 801
\bibitem[\protect\citeauthoryear{Balbus
\& Papaloizou}{1999}]{balbus99} Balbus S.~A., Papaloizou J.~C.~B., 1999, ApJ, 521, 650
\bibitem[\protect\citeauthoryear{Bildsten et
al.}{1997}]{bildsten97} Bildsten L., et al., 1997, ApJS, 113, 367
\bibitem[\protect\citeauthoryear{Bradt, Rothschild,
\& Swank}{1993}]{rxte} Bradt H.~V., Rothschild R.~E., Swank J.~H., 1993, A\&AS, 97, 355
\bibitem[\protect\citeauthoryear{Brandenburg et
al.}{1995}]{brandenburg95} Brandenburg A., Nordlund A., Stein R.~F.,
Torkelsson U., 1995, ApJ, 446, 741
\bibitem[\protect\citeauthoryear{Caballero et
al.}{2008}]{caballero08} Caballero I., et al., 2008, A\&A, 480, L17
\bibitem[\protect\citeauthoryear{Churazov, Gilfanov,
\& Revnivtsev}{2001}]{churazov01} Churazov E., Gilfanov M., Revnivtsev M., 2001, MNRAS, 321, 759
\bibitem[\protect\citeauthoryear{Corbet}{1984}]{corbet84} Corbet R.~H.~D., 1984, A\&A, 141, 91
\bibitem[\protect\citeauthoryear{Davidson
\& Ostriker}{1973}]{davidson73} Davidson K., Ostriker J.~P., 1973, ApJ, 179, 585
\bibitem[\protect\citeauthoryear{Finger, Wilson,
\& Harmon}{1996}]{finger96} Finger M.~H., Wilson R.~B., Harmon B.~A., 1996, ApJ, 459, 288
\bibitem[\protect\citeauthoryear{Ghosh
\& Lamb}{1979}]{gl79} Ghosh P., Lamb F.~K., 1979, ApJ, 234, 296
\bibitem[\protect\citeauthoryear{Gilfanov
\& Arefiev}{2005}]{gilfanov05} Gilfanov M., Arefiev V., 2005, astro, arXiv:astro-ph/0501215
\bibitem[\protect\citeauthoryear{Hirose, Krolik,
\& Stone}{2006}]{hirose06} Hirose S., Krolik J.~H., Stone J.~M., 2006, ApJ, 640, 901
\bibitem[\protect\citeauthoryear{Hoshino
\& Takeshima}{1993}]{hoshino93} Hoshino M., Takeshima T., 1993, ApJ, 411, L79
\bibitem[\protect\citeauthoryear{Kotov, Churazov,
\& Gilfanov}{2001}]{kotov01} Kotov O., Churazov E., Gilfanov M., 2001, MNRAS, 327, 799
\bibitem[\protect\citeauthoryear{Kraicheva et
al.}{1999}]{kraicheva99} Kraicheva Z., Stanishev V., Genkov V., Iliev L., 1999, A\&A, 351, 607
\bibitem[\protect\citeauthoryear{Lamb, Pethick,
\& Pines}{1973}]{lamb73} Lamb F.~K., Pethick C.~J., Pines D., 1973, ApJ, 184, 271
\bibitem[\protect\citeauthoryear{Lipunov
\& Shakura}{1976}]{lipunov76} Lipunov V.~M., Shakura N.~I., 1976, SvAL, 2, 133
\bibitem[\protect\citeauthoryear{Lyubarskii}{1997}]{lyubarskii97}
Lyubarskii Y.~E., 1997, MNRAS, 292, 679
\bibitem[\protect\citeauthoryear{Matthaeus et al.}{1991}]{matthaeus91}
Matthaeus W.H, Klein L.W., Ghosh S., Brown M.R., 1991, J. Geophys. Res., 96, 5421
\bibitem[\protect\citeauthoryear{Oda et al.}{1974}]{oda74}
Oda M., Takagishi K., Matsuoka M., Miyamoto S., Ogawara Y., 1974, PASJ, 26,
303
\bibitem[\protect\citeauthoryear{Pandel, C{\'o}rdova,
\& Howell}{2003}]{pandel03} Pandel D., C{\'o}rdova F.~A., Howell S.~B., 2003, MNRAS, 346, 1231
\bibitem[\protect\citeauthoryear{Postnov et
al.}{2008}]{postnov08} Postnov K., Staubert R., Santangelo A., Klochkov D., Kretschmar P., Caballero I., 2008, A\&A, 480, L21
\bibitem[\protect\citeauthoryear{Pringle\& Rees}{1972}]{pringle72} Pringle J.~E., Rees M.~J., 1972, A\&A, 21, 1
\bibitem[\protect\citeauthoryear{Rappaport, Doxsey,
\& Zaumen}{1971}]{rappaport71} Rappaport S., Doxsey R., Zaumen W., 1971, ApJ, 168, L43
\bibitem[\protect\citeauthoryear{Revnivtsev et
al.}{2006}]{revnivtsev06} Revnivtsev M., et al., 2006, A\&A, 447, 545
\bibitem[\protect\citeauthoryear{Shakura}{1975}]{shakura75}
Shakura N.~I., 1975, SvAL, 1, 223
\bibitem[\protect\citeauthoryear{Terrell}{1972}]{terrell72}
Terrell N.~J.~J., 1972, ApJ, 174, L35
\bibitem[\protect\citeauthoryear{Tsygankov et
al.}{2007}]{tsygankov07} Tsygankov S.~S., Lutovinov A.~A., Churazov
E.~M., Sunyaev R.~A., 2007, AstL, 33, 368
\bibitem[\protect\citeauthoryear{Uttley
\& McHardy}{2001}]{uttley01} Uttley P., McHardy I.~M., 2001, MNRAS, 323, L26
\bibitem[\protect\citeauthoryear{Wijnands
\& van der Klis}{1999}]{wijnands99} Wijnands R., van der Klis M., 1999, ApJ, 514, 939
\bibitem[\protect\citeauthoryear{Ziolkowski}{1985}]{ziolkowski85}
Ziolkowski J., 1985, AcA, 35, 185
\end{thebibliography}
\end{document}